\documentclass[runningheads]{llncs}
\usepackage{xcolor}

\usepackage{dirtree}
\usepackage{graphicx}
\usepackage{url}
\usepackage{textcomp}
\usepackage{listings}

\lstdefinestyle{BashInputStyle}{
  language=bash,
  upquote=true,
  breaklines=true,
  breakatwhitespace=true,
  basicstyle=\small\ttfamily,
  frame=tblr,
  aboveskip=10pt,
  belowskip=10pt,
  showstringspaces=false,
  linewidth=0.9\linewidth,
  xleftmargin=0.1\linewidth
}

\begin{document}
\title{A User-oriented Portable, Reproducible, and Scalable Software Ecosystem}
\titlerunning{A User-oriented Software Ecosystem}
%
\author{Alfio~Lazzaro\inst{1}\orcidID{0000-0003-4256-8270} \and
Utz-Uwe~Haus\inst{1}$^{*}$\orcidID{0000-0001-7292-9984} \and\\
Sandrine~Charousset\inst{2} \and
Nina~Mujkanovic\inst{1}
}
\authorrunning{A.~Lazzaro et al.}
\institute{HPE HPC/AI EMEA Research Lab, 4051 Basel, Switzerland
\email{\{alfio.lazzaro,utz-uwe.haus,nina.mujkanovic\}@hpe.com}
\and
EDF Lab Paris Saclay, 91120 Palaiseau, France\\
\email{sandrine.charousset@edf.fr} \\
$^{*}$~\emph{Corresponding author}
}
\maketitle
%
\begin{abstract}
It is normal for scientists to perform their research on a diverse set of hardware, ranging from laptops and workstations to supercomputers and cloud resources.
The standard scenario requires a mix of these resources.
In this paper we describe a software ecosystem that enables users to rely on the same development environment for running their workflows across the different computational resources.
We describe a modular, unified command-line interface that allows for the interaction with a user-workflow across diverse hardware platform. 
The software ecosystem has been successfully tested as part of the \textsc{plan4res} EU H2020 project. 
It can be extended to other projects with similar requirements, so that they can benefit from the same approach for executing computational workflows.

\end{abstract}
\section{Introduction}
\label{sec:introduction}

Scientists typical have access to a range of computing hardware for their research, from laptops and workstations (\emph{local} resources), to supercomputers and cloud resources (\emph{remote} resources). A standard scenario requires a mix of these hardware resources, as science projects may consist of numerous steps and methods involving multiple applications and running on various devices, from classical high performance computing (HPC) systems to cloud-based services. For example, users may carry out fast prototyping of their workflows on local resources at their disposal, then move to HPC remote resources for specific production workflows requiring more performant or specialized hardware (e.g. fast interconnects for parallel job executions, or GPUs for computation). 

We note that users often prefer their local systems due to familiarity with specific operating systems (OSs), e.g. \textsc{Windows} or \textsc{macOS}, or the availability of applications not typically available on remote resources which often have a \textsc{Linux} system installed, possibly with limited privileges.
Pre- and post-data processing, for example, can require custom or commercial applications, where the data cannot be shared outside the user's institution due to security reasons.
Conversely, data may instead be located on the remote site, and only a portion of it needed for user tests on the local machines, leaving the data closer to the computation (\emph{edge} computing).

 Cloud computing technologies have gained traction in HPC for their benefits, including resource dynamism compared to the more static approach of traditional workload managers such as Slurm and PBS, workload automation, reproducibility via virtualization, and resilience owing to their microservices approach. Supercomputers offer access to advanced HPC computing techniques and massive processing capability for grand challenges and scientific discovery. In recent years we have seen the convergence between the two spheres, as typical HPC applications such as MPI have crept into the cloud ecosystem, and HPC system vendors and sites have incorporated the notion of microservices and containerization. An overview on the differences and commonalities between cloud and supercomputers, and how to run workflows on each, can be found elsewhere~\cite{hpc_cloud1,hpc_cloud3,hpc_cloud2,hpc_cloud4,nina_tiziano_sc}.
 
 This convergence and the commoditization of HPC on one side, and the growing complexity of simulation and data analysis workflows on the other, have made workflow management a necessity. The need for the efficient use of resources, as well as the increased requirements to provide constant, reliable, and reproducible results, has driven the creation of a multitude of workflow management applications, which are often domain dependant~\cite{nina_tiziano_sc}. We, instead, envision a software ecosystem that enables users to rely on the same development environment (same \emph{user-experience}) for running their workflows across local and remote resources. 

In this paper, we describe a modular, unified command-line interface using containerization to enable the launching of jobs via a workload manager on a mix of hardware resources. In Section~\ref{sec:requirements}, we describe the requirements that have driven the software ecosystem development, based upon the domain-expert user's perspective (\emph{user-oriented}). As users can integrate their specific applications in the software ecosystem as part of their favourite development workflow environment, and run seamlessly on local and remote resources, we consider this approach \emph{portable} and  \emph{reproducible}, and it can interface with a \emph{scalable} workload manager to carry out complex workflows. The implementation details of the software ecosystem are given in Section~\ref{sec:implementation}. The ecosystem was developed as part of the \textsc{plan4res} EU H2020 project~\cite{CharoussetBrignolvanAckooijOudjaneetal.2021}, and we present a workflow application use case in Section~\ref{sec:plan4res}. Finally, concluding remarks and an outlook are given.

\subsection{Related Work}

Various frameworks have emerged to support user workflows from simple desktop calculations to complex activities that require large infrastructure shared by a vast community of users. Science gateways combine computational resources from grid, cloud, and supercomputers, while reducing the learning curve and barriers to entry by implementing Web and mobile applications~\cite{science_gateways}. Jupyter Notebook has proven convenient for running workflows via a Web interface on local and remote systems. Colonnelli \emph{et al.} have shown a way to execute complex distributed workflows with Jupyter, with a unified interface to Cloud and HPC for scientific applications~\cite{COLONNELLI2022282}. Although these methods provide user-friendly Web interfaces out-of-the-box, they are mostly suited for scripting language applications, and not really meant for HPC execution as they lack sufficient support for the version control required for collaborative activities~\cite{10.1093/gigascience/giad113}.

The \textsc{Emerald} system allows to build workflows to be executed between local resources and cloud, where users can decide which computation steps should run on remote resources~\cite{Qian2016}. An abundance of studies concerning software environments that allow portable, reproducible, and scalable deployments, targeting cloud and HPC resources, exists, see for example Refs.~\cite{10.1145/3311790.3396659,10.7554/eLife.71069}. Their common approach is to use software containers as execution building-blocks. We follow a similar approach, however we also provide a unified command-line interface enabling the launching of jobs via a workload manager on a diverse set of hardware and OSs.

\section{Software Ecosystem Requirements}
\label{sec:requirements}

\begin{figure}[t]
\centering
\includegraphics[width=.55\textwidth]{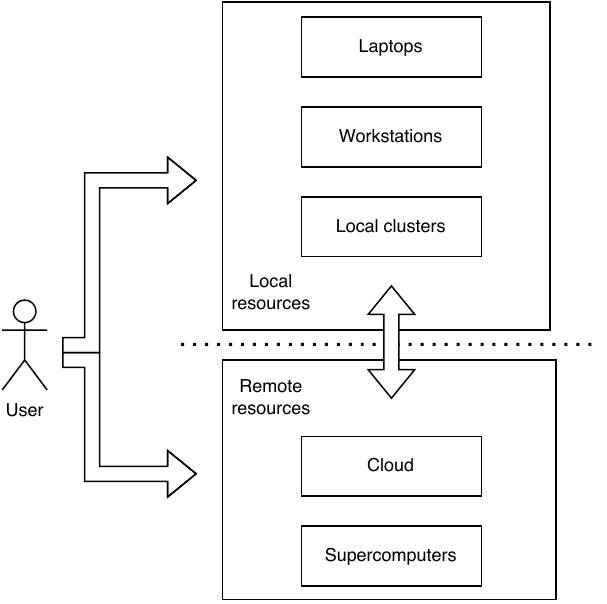}
\caption{User-oriented approach for running on local and remote computational resources.} 
\label{fig:local_remote}
\end{figure}

With reference to Section~\ref{sec:introduction}, we aim to design a software ecosystem that allows users to run on different computing resources with minimal burden. Here we distinguish between local and remote resources, as show in Fig.~\ref{fig:local_remote}. In particular, local resources can have different OSs (\textsc{Windows}, \textsc{macOS}, \textsc{Linux}), while we assume that remote resources are only based on \textsc{Linux} for intensive and scalable computational tasks. Therefore, the goal is to have a single entry-point for the execution control of the hardware resources that integrates the local software environment, with the option of specific configurations depending on which resource is used.

We base the design requirements of the software ecosystem on a user-oriented approach, where we generalize and abstract the domain-experts software for specific hardware. The requirements were introduced during the \textsc{plan4res} project, and generalized later on with the introduction of additional requirements, resulting in the following:
\begin{itemize}
    \item unified control of the resources via a shell command-line interface (CLI);
    \item support for multiple OSs and compute systems;
    \item a minimal set of commands (possibly one) to interact with the resources to ensure good usability for domain-expert (non-technical) users;
    \item possibility to enable specific configurations via files or environment variables;
    \item common pre-installed components and a mechanism to add extra software to the platform as needed;
    \item relocation of the environment across systems with a simple drop-in replacement procedure, i.e. no installation needed and binary exchangeability;
    \item presumed multi-user shared environment;
    \item possibility to define and execute data-driven, complex workflows composed of Shell- and Python-scripts, as well as binary executables, including data transfer out of and into the storage platform.
\end{itemize}
The technical implementation to address these requirements is described in Section~\ref{sec:implementation}.

\section{Software Ecosystem Implementation}
\label{sec:implementation}

The fundamental assumption for the software environment implementation is that all platform tools will run in a modern \textsc{Linux}-based software environment, on hosts with \texttt{x86\_64} CPUs, where we avoid requiring \textsc{Linux}-specificity by relying on topic-specific standards only, in particular \textsc{POSIX}. The user CLI must run in a terminal shell, which is available by default on \textsc{Linux} and \textsc{macOS} systems, while for \textsc{Windows} users we suggest to use the \textsc{git-bash} shell~\footnote{\url{https://www.atlassian.com/git/tutorials/git-bash}.}. The scripts, data, configuration files, and application executables of the software ecosystem are organized in a specific directory structure, as described in Section~\ref{sec:directory_structure}. 

To ease deployment, we create a containerized environment\index{container} based
on the \textsc{Singularity} container infrastructure~\cite{singularity_pub}.
This container acts as~\emph{executor}, meaning that all tools are based in and operated by it. The use of containers has been proven to be a very convenient way to abstract from the underlying host OS, permitting software to run reliably when moved from one computing environment to another (assuming hardware compatibility), without having to build and configure separately~\cite{merkel2014docker}. This fulfills the requirement of running on a variety of hardware resources of interest. 

While the \textsc{Singularity} container can run natively on a \textsc{Linux} OS host, it cannot on \textsc{Windows} or \textsc{macOS} OSs~\footnote{ \url{https://docs.sylabs.io/guides/3.8/admin-guide/installation.html#installation-on-windows-or-mac}.}, which is one of our requirements. For this reason, a minimal installation of a \textsc{Linux} virtual machine (VM) is required. We again assume that hosts have \texttt{x86\_64} CPUs. We describe the container and the VM configurations in Sections~\ref{sec:executor} , as well as components required pre-installed, and a mechanism to add extra software to the platform as needed. Finally, in Section~\ref{sec:workflow} we address the implementation requirements to define and execute data-driven, complex workflows.

\subsection{Directory Structure}
\label{sec:directory_structure}

\begin{figure}[t]
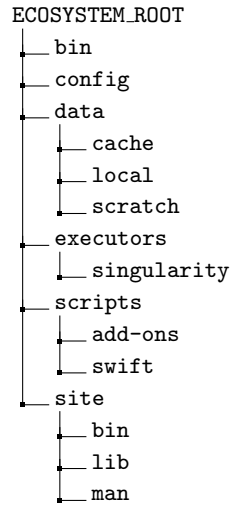

    \centering
    \begin{minipage}{.3\textwidth}
    \dirtree{%
    .1 ECOSYSTEM\_ROOT. 
     .2 bin.
     .2 config.
     .2 data.
      .3 cache.
      .3 local.
      .3 scratch.
     .2 executors.
      .3 singularity.
     .2 scripts.
      .3 add-ons.
      .3 swift.
    .2 site.
      .3 bin.
      .3 lib.
      .3 man.
    }
    \end{minipage}
    \caption{Software ecosystem directory structure.}
    \label{fig:dirtree}
\end{figure}

The software ecosystem is organized in a directory structure. The root directory, defined by the user on the computer system, needs to
\begin{itemize}
\item be accessible from all compute nodes that will concurrently work on
  the data;
\item have enough space for all data and the software environment;
\item satisfy \textsc{POSIX} file system semantics.
\end{itemize}
We refer to this directory as 
\texttt{ECOSYSTEM\_ROOT}. This is also the name of an environment variable that contains the directory name, which tools can query at run-time. Inside this root directory, the directory structure shown in Fig.~\ref{fig:dirtree} is mandated, and can be relied upon when writing programs for use within
the framework.

The \texttt{ECOSYSTEM\_ROOT/bin} directory contains all software that is needed for workflow execution and must be in the user's search \texttt{PATH} so that the operating system can find it. This software will run the specific commands via the container available in the \texttt{ECOSYSTEM\_ROOT/executors} directory, which includes the necessary files for building it.

The \texttt{ECOSYSTEM\_ROOT/config} directory contains the configuration file(s), see Section~\ref{sec:workflow}.

All data will reside below the \texttt{ECOSYSTEM\_ROOT/data} directory. This directory takes the role of the staging area and also the results storage area. Data cached from external sources will be stored below \texttt{ECOSYSTEM\_ROOT/data/cache}. Local data that is specific to the user's workspace, in particular private data, should be stored below \texttt{ECOSYSTEM\_ROOT/data/local}. The \texttt{ECOSYSTEM\_ROOT/data/scratch} directory is used as the working directory of tools and can be used for volatile data during program runs. 

The \texttt{ECOSYSTEM\_ROOT/scripts} directory can contain more complex scripts to execute repeated tasks using executables from \texttt{ECOSYSTEM\_ROOT/bin}.

The \texttt{ECOSYSTEM\_ROOT/site} can be used by users to install their own local software (comparable to \texttt{/usr/local} on Unix systems), which already guarantees that
\texttt{ECOSYSTEM\_ROOT/site/bin} is in \texttt{PATH}, \texttt{ECOSYSTEM\_ROOT/site/lib} is in \texttt{LD\_LIBRARY\_PATH}, and \texttt{ECOSYSTEM\_ROOT/site/man} is in the \texttt{MANPATH}.

The directory structure can be automatically created by cloning a
\texttt{GIT} repository, for which the created directory represents the \texttt{ECOSYSTEM\_ROOT} path. This repository contains all scripts needed for the environment and can be shared between multiple users via the standard \texttt{GIT} features, like branching and forking.

\subsection{Executor environment}
\label{sec:executor}

The execution in the platform is triggered by calling a \textsc{BASH}  script \texttt{driver} in the
\texttt{ECOSYSTEM\_ROOT/bin} directory, which acts as the only \emph{entry-point} for all users activities to be executed in the software ecosystem, including the execution on the container image with arguments designating the tool to execute. This command will parse the configuration files under \texttt{ECOSYSTEM\_ROOT/config} and modify the environment as needed, then start the appropriate script or binary within the container environment. The \texttt{driver} command acts like a special shell; in fact, running it without arguments will start a shell in the container
environment with all paths set up appropriately so that scripts and
tools are found and can be executed:\\
\begin{minipage}{\linewidth}
\begin{lstlisting}[style=BashInputStyle]
$ ECOSYSTEM_ROOT/bin/driver
;; Run 'driver -h' to get help.
[ENV] ~ > 
\end{lstlisting}
\end{minipage}\\
Alternatively, the command to be executed (with its arguments) can be directly passed during the \texttt{driver} invocation.
The configuration files are \textsc{BASH}
shell-script fragments with default environment variables set suitable for most users, depending on the specific workflow. The users can change the values of these variables by editing the configuration files, or by exporting the values directly in the environment. For this reason, the variables are set via the syntax \texttt{VARIABLE\_NAME=\$\{VARIABLE\_NAME:-DEFAULT\_VALUE\}}. For example, there can be multiple containers under the \texttt{ECOSYSTEM\_ROOT/executors} directory that can be selected via a variable in the configuration file.

The minimal required dependencies for a \textsc{Linux} host system are: the \textsc{BASH} shell, the \textsc{GIT} command, and the \textsc{Singularity} application (we use version 3.11). We test on \textsc{Debian~10}, \textsc{Ubuntu~20.04}, \textsc{Ubuntu~22.04}, \textsc{Fedora~35}, and \textsc{SUSE Linux Enterprise Server 15}. The container image binds the host \texttt{ECOSYSTEM\_ROOT} directories tree (read and write access), so that it can access all files of the ecosystem of the host system.

For the \textsc{Windows} and \textsc{macOS} host systems, we based our implementation on \textsc{Vagrant}~\footnote{\url{https://developer.hashicorp.com/vagrant}.} and \textsc{VirtualBox}~\footnote{\url{https://www.virtualbox.org/}.} to conveniently set up a suitable VM. The VM is based on the \textsc{Debian~10} distribution. A \texttt{Vagrantfile} in the \linebreak
\texttt{ECOSYSTEM\_ROOT} directory is provided for the installation of the dependencies. The user starts the VM when running the \texttt{driver} command (the \texttt{driver} will check if the VM is running, otherwise it will start it via the \texttt{vagrant up} command). Therefore there is no direct connection to the VM, making all ecosystem CLI commands independent of the host OS. We test on \textsc{Windows~10~21H2} and \textsc{macOS~11} and 12. The VM introduces an intermediate layer between the host system and the container image running on it. For this reason, the VM has to mount the host \texttt{ECOSYSTEM\_ROOT} directories tree (described in the \texttt{Vagrantfile} and done during the provisioning of the \textsc{Vagrant} VM), so that the container image can bind it.
The \texttt{driver} itself will deal with the intermediate layer, so that the users will not see any difference with respect to running directly on a \textsc{Linux} host system. In conclusion, the entire
\texttt{ECOSYSTEM\_ROOT} directory tree is shared (for reading and
writing) between the Host (\textsc{Linux}, \textsc{Windows}, \textsc{macOS}) and the container.

The users can build the container locally via a script available in the \linebreak 
\texttt{ECOSYSTEM\_ROOT/bin} directory, or download a pre-built version available from a given URL. Note that the build procedure will directly occur on the \textsc{Linux} host or in the VM for \textsc{Windows} and \textsc{macOS} hosts. The containers include all the required packages for building the users applications such as compilers, build tools (e.g. \texttt{cmake}), Python modules, parallel libraries (e.g. \texttt{MPI}), and scientific libraries. Therefore, the container serves as a common working platform for the users, but by design does not contain the users applications. We implement the building of the pre-built container as part of a continuous-integration (CI) procedure linked to the
the \textsc{GIT} repository of the software ecosystem. Any change of the files used for building the container pushed to the \textsc{GIT} repository will trigger the CI for rebuilding the container that is then stored as an artifact. In this way users can download this image and do not need to build it on their local systems.
By default, the users download the containers, unless they have built a local container and set to use it in the configuration file. Then, for every execution of the \texttt{driver} there will be a check whether the local
cached copy of the container image matches the remote one, and updates will be downloaded automatically. This check can be avoided via a configuration setting to preserve the local cached copy.

User specific software can be installed using a recipe-based \emph{add-on} installation infrastructure. 
These software packages will be built and executed via the container. This procedure allows to install tools that can not be directly inserted into the container (mainly due to licensing issues). The advantage of having the add-ons not be part of the container is that users can update the container with new functionalities without the need of reinstalling them. An add-on requires a recipe file, stored under the \texttt{ECOSYSTEM\_ROOT/scripts/add-ons}
directory. The recipe is an executable script, based on \texttt{makefile} syntax, that supports a common set of targets, in particular \texttt{install}, \texttt{update}, \texttt{status}, \texttt{uninstall}, and \texttt{help}. The add-ons are installed into the directory \linebreak
\texttt{ECOSYSTEM\_ROOT/scripts/add-ons/install} so that they can be executed via the \texttt{driver}. Binaries, libraries, and man pages are automatically added to the corresponding environment variables of the container. A user can then query the list of available add-ons via the command:\\
\begin{minipage}{\linewidth}
\begin{lstlisting}[style=BashInputStyle]
$ ECOSYSTEM_ROOT/bin/driver add-on
<add-on name> : <installed | not installed>

Use 'add-on <add-on name>' to install an add-on.
Use 'add-on <add-on name> help' to see a list of specific options per each add-on.
\end{lstlisting}
\end{minipage}\\
The add-ons can be parallel MPI applications, which the \texttt{driver} will launch through various mechanisms, such as \texttt{mpiexec/mpirun} commands or the \texttt{srun} command for the Slurm batch system.

Finally, we provide a script that can archive the entire ecosystem installed software so that it can be relocated to another system, including the possibility to store backups, or alternatively synchronize (via the \texttt{rsync} command) files between two locations, possibly between different users.

\subsection{Workflow Execution}
\label{sec:workflow}

\begin{figure}[t]
\centering
\includegraphics[width=.6\textwidth]{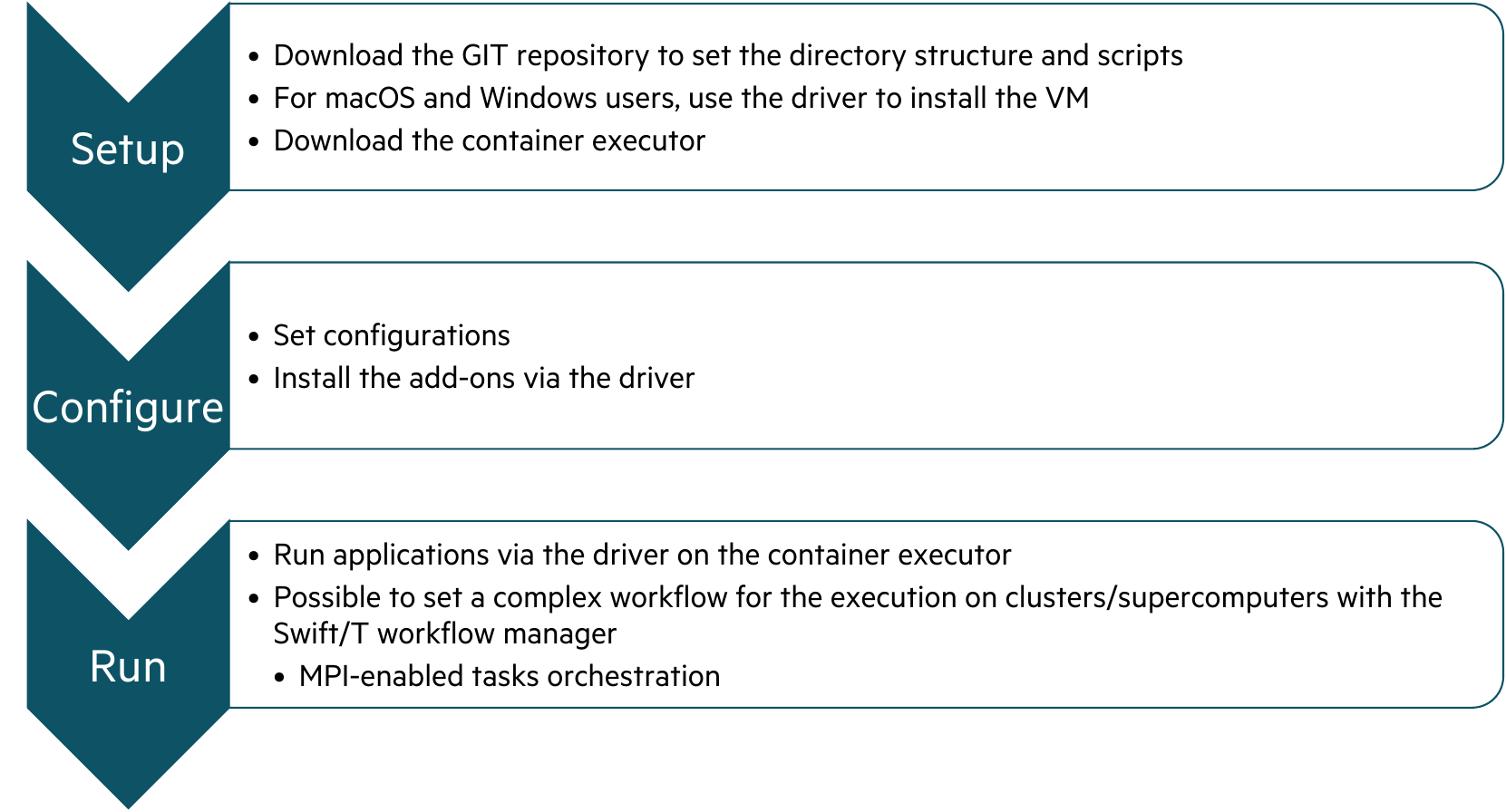}
\caption{Schematic example of the software ecosystem installation and usage.} 
\label{fig:steps}
\end{figure}

In this section we describe a possible user installation workflow that employs the software ecosystem described in this paper. There are 3 steps, as shown in Fig.~\ref{fig:steps}: setup, configure, and run. We assume that all required dependencies mentioned at the beginning of Section~\ref{sec:implementation} are installed on the host system. As a result of our abstraction, the same procedure is valid for local and remote resources, independently of the OS used for the local resources, with the possibility to relocate the software ecosystem across systems.

The first step (setup) is to clone the \texttt{GIT} repository which provides the directories tree and the default files, after which users can decide to use the remote container, which will be downloaded by calling the command:\\
\begin{minipage}{\linewidth}
\begin{lstlisting}[style=BashInputStyle]
$ ECOSYSTEM_ROOT/bin/driver -t
\end{lstlisting}
\end{minipage}\\
This command will download and cache the container file, and eventually test it via the installation integrity tests provided in the container. Note that every call to the \texttt{driver} command will trigger a check whether the local cached copy of the \textsc{Singularity} image matches the one on the remote site, and will download updates automatically, unless the user enables the configuration variable to preserve the local copy.

The next step (configure) is to configure and install the proper add-ons. We again use the \texttt{driver} command, as described at the end of  Section~\ref{sec:executor}.
Eventually, we can use the installed commands via the \texttt{driver} command, as described at the beginning of Section~\ref{sec:executor}.

Finally, the user can describe complex workflows with correlated data dependencies and executions to be performed (run step). The components of the overall workflow are conceptually coupled and interact via data transfers. While it is possible to execute a complete
run of the tools by executing every task individually, the true power stems from the integration of all steps into a reproducible workflow described in a documented, machine-executable workflow description. There are two approaches:
\begin{itemize}
    \item \textbf{Manual workflows} can be constructed by explicitly using data files and scripting in a common scripting language like Shell- and Python-scripts. The users can run these scripts via the usual \texttt{driver} command.
    \item \textbf{Automated workflows} are defined using \textsc{Swift/T}, an MPI-oriented workflow language and runtime system~\cite{swift-t:13}. Swift/T is designed to enable the execution of vast numbers of very small tasks across an MPI-enabled computing system. The tasks could be composed of Shell- and Python-scripts, as well as binary executables, all installed through the add-on procedure. This approach permits an execution model where decisions about the task execution are driven by the availability of their required input data, and where data transfers can be performed by the tool in an efficient manner adapted to the capabilities of the system, instead of being planned by the user.
\end{itemize}

\section{\textsc{Plan4res} Use-case}
\label{sec:plan4res}

The \textsc{plan4res} (P4R) EU H2020 project aimed to provide a tool for optimizing and simulating the European electricity system, validating the results through real size case studies~\cite{CharoussetBrignolvanAckooijOudjaneetal.2021}. Operating the electricity system with the 2050 targeted shares of Renewable Energy Systems (RES) will only be possible and affordable if both grids and generation assets evolve towards a system designed to maximise its capacity to host such amounts of RES. This requires optimizing existing assets and new investments, while making the best use of all flexibilities, such as controllable power plants, but also storage, use of interconnections, and demand control.  To achieve this, an integrated representation of the system becomes necessary, involving overcoming significant technical hurdles in the implementation and maintenance of complex models with several nested layers of structure and general and flexible algorithms for solving them, thus leading to problem sizes that grow tremendously with each level of detail modeled.

The P4R environment has been developed following the software ecosystem implementation described in this paper~\cite{p4r-env}. The main \textsc{GIT} repository is available on \textsc{GitLab}~\footnote{\url{https://gitlab.com/cerl/plan4res/p4r-env}.}. The container executor is presented as a \textsc{GIT} submodule under the \texttt{executors/singularity} directory, which points to another \textsc{GitLab} repository~\footnote{\url{https://gitlab.com/cerl/plan4res/p4r-exec-singularity}.}. 
This solution allows to separate the container development versus the main P4R ecosystem. We use the  \textsc{GitLab} CI to build the container, which is then stored as an artifact of the repository. For convenience, the \texttt{driver} has been renamed to \texttt{p4r}. Two versions of the container are provided, containing MPICH and OpenMPI implementations, respectively. Add-ons provide users access to mathematical optimization packages~\cite{p4r-env}. The models and the solution algorithms have been implemented using the open-source \emph{Structured Modeling System++} (\texttt{SMS++})~\footnote{\url{https://smspp.gitlab.io/}.} package, which is provided as an add-on. 

\begin{figure}[t]
\centering
\includegraphics[width=.6\textwidth]{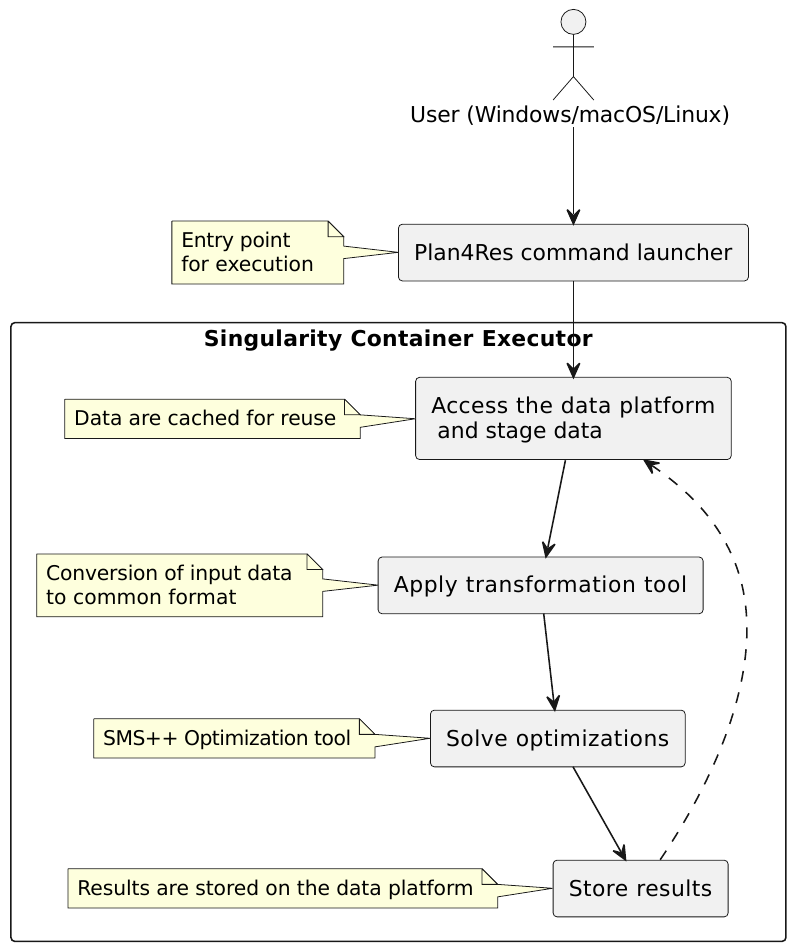}
\caption{\textsc{Plan4res} schematic workflow.} 
\label{fig:p4r_workflow}
\end{figure}

The full P4R workflow consists of the following base steps: 
\begin{itemize}
    \item access and stage data from an external source;
    \item run a transformation tool (a Python script, provided as add-on) on the data to convert them to a format for the following computation;
    \item solve the optimizations via \texttt{SMS++};
    \item store the results of the optimizations.
\end{itemize}
A diagram of the P4R workflow is shown in Fig.~\ref{fig:p4r_workflow}. 
The workflow has been deployed both in a manual as well as an automated workflow and successfully executed on \textsc{plan4res} partner's systems: laptops, private clusters, HPE Cray EX Supercomputers, and an AWS ParallelCluster instance~\footnote{\url{https://docs.aws.amazon.com/parallelcluster/}.}.
Therefore, the P4R software ecosystem has been successful for providing a fast user prototyping and deployment environment for workflow execution on different hardware resources.

\section{Conclusion}

The appearance of and convergence between such varied compute sources as HPC, cloud technologies, edge computing, novel hardware including GPUs and FPGAs in the recent years has created both opportunities as well as challenges. Users now have a more flexible, portable, and easy to deploy option via containerization, which renders the sharing and cooperation of research easier, but also introduces complexity in the workflow. In this paper we presented a software ecosystem that offers users a modular, unified command-line interface to enable the same development environment whether running their workflows on local or remote resources, thus allowing for easy deployment, letting users focus on their research instead of technological implementation specificities.

The software ecosystem has been successfully tested as part of the \textsc{plan4res} EU H2020 project, and was used to assess the feasibility and cost of a recently published, long-term energy scenario from the \textsc{openENTRANCE} EU H2020 project~\cite{charousset_2023_8288993}. Currently, it is used within the 
\textsc{openMod4Africa}~\footnote{\url{https://openmod4africa.eu/}.} project for case studies in Western Africa (both at the level of the whole West Africa Power Pool region, and at the level of Senegal) and Eastern Africa (for a case study at the level of the whole East Africa Power Pool and a study focused on Ethiopia). The case studies will be conducted by African experts from universities and the power pools, with the help of the European experts bringing the models. It is expected that the software ecosystem will be used by more than ten African institutions in the next two years. Another European project which will use the software ecosystem is MANOEUVRE~\footnote{https://man0euvre.eu/}. This project has just started and is dedicated to designing transition scenarios for Europe as well as defining methodologies and providing tools for conducting the national exercises aiming at developing the National Energy and Climate Plans. 

Our software ecosystem can be extended to other projects with similar requirements, such that they can benefit from the same flexible approach for executing computational workflows.


\begin{credits}
\subsubsection{\ackname} The \textsc{plan4res} project received funding from the European Union's Horizon 2020 research and innovation programme under grant agreement No 773897.

\subsubsection{\discintname}
The authors have no competing interests to declare that are relevant to the content of this article.
\end{credits}

%
%
\bibliographystyle{splncs04}
\bibliography{biblio}

\end{document}